\documentclass[12pt]{article}
\begin{document}
\title{\bf Energy-Momentum of the Friedmann Models in General Relativity
and Teleparallel Theory of Gravity}

\author{M. Sharif \thanks{msharif@math.pu.edu.pk} and M. Jamil Amir
\thanks{mjamil.dgk@gmail.com}\\
Department of Mathematics, University of the Punjab,\\
Quaid-e-Azam Campus, Lahore-54590, Pakistan.}

\date{}

\maketitle

\begin{abstract}
This paper is devoted to the evaluation of the energy-momentum
density components for the Friedmann models. For this purpose, we
have used M${\o}$ller's pseudotensor prescription in General
Relativity and a certain energy-momentum density developed from his
teleparallel formulation. It is shown that the energy density of the
closed Friedmann universe vanishes on the spherical shell at the
radius $\rho=2\sqrt{3}$. This coincides with the earlier results
available in the literature. We also discuss the energy of the flat
and open models. A comparison shows a partial consistency between
the M${\o}$ller's pseudotensor for General Relativity and
teleparallel theory. Further, it is shown that the results are
independent of the free dimensionless coupling constant of the
teleparallel gravity.
\end{abstract}


\section{Introduction}

The localization of energy and momentum [1] in General Relativity
(GR) is an open, most challenging and controversial problem. In the
framework of GR, many attempts have been made to calculate energy
distribution by using different approaches. The use of
energy-momentum complexes for the localization of energy and
momentum is one of these methods. Many physicists, such as Einstein
[2], Landau-Lifshitz [3], Papapetrou [4], Bergmann [5], Tolman [6],
Weinberg [7] and M${\o}$ller [8] have given their own definitions
for the energy-momentum complex. Most of these are coordinate
dependent while the M${\o}$ller expression gives an energy value
independent of the choice of spatial coordinates; this is not the
case for the momentum.

The lack of a generally accepted definition of energy-momentum in a
curved spacetime has led to doubts regarding the idea of energy
localization. According to Misner et al. [1], energy is localizable
only for spherical systems. Cooperstock and Sarracino [9] came up
with the view that if energy is localizable only for spherical
systems, then it can be localized in any system. Bondi [10] argued
that a non-localizable form of energy is not allowed in GR. After
this, an alternative concept of energy, called quasi-local energy,
was developed. The use of quasi-local masses to obtain
energy-momentum in a curved spacetime do not restrict one to use
particular coordinate system. A large number of definitions of
quasi-local masses have been proposed, those by Penrose and many
others [11-13]. An excellent review article about quasi-local
energy-momentum and angular momentum in GR has been given by
Szabados [14]. Although these quasi-local masses are conceptually
very important, yet these definitions have some problems. Bergqvist
[15] considered seven different definitions of quasi-local masses
and computed them for Reissner-Nordstrom and Kerr spacetimes. He
concluded that no two of the seven definitions provided the same
result. The seminal concept of quasi-local masses of Penrose cannot
be used to handle even the Kerr metric [16]. The present quasi-local
mass definitions still have inadequacies.

It is believed that different energy-momentum distribution would be
obtained from different energy-momentum complexes. Virbhadra [17,18]
revived the interest in this approach. He and his co-workers [18-22]
considered many asymptotically flat spacetimes and showed that
several energy-momentum complexes could give the same result for a
given spacetime. They also carried out calculations in a few
asymptotically non-flat spacetimes using different energy-momentum
complexes and found encouraging results. Aguirregabiria at el. [23]
proved that several energy-momentum complexes can provide the same
results for any Kerr-Schild class metric. Xulu [24,25] extended this
investigation and found that Melvin magnetic universe and Bianchi
type I universe provided the same energy distribution. One of the
authors [26] found several examples which did not provide the same
result in all prescriptions. Chang at el. [27] showed that every
energy-momentum complex could be associated with a particular
Hamiltonian boundary term. Therefore, the energy-momentum complexes
may also be considered as quasi-local. According to the Hamiltonian
approach, the various energy-momentum expressions are each
associated with distinct boundary conditions [27,28].

The beauty and speciality of the M${\o}$ller prescription is that
one can use any spatial coordinate system for evaluating the energy,
while the other prescriptions restrict one to use the Cartesian
coordinate system only for obtaining meaningful results. On the
basis of this fact, Lessner [29] concluded that \textit{M${\o}$ller
definition is a powerful concept of energy and momentum in GR}.
Also, the literature [30-33] shows that the M${\o}$ller
energy-momentum complex is a good tool for evaluating energy
distribution in a given spacetime. Thus the use of M${\o}$ller
prescription looks more interesting, useful and appropriate while
finding the energy distribution. The results obtained in
[18,20,22,25,26,33] indicate that the energy distribution is
different for some particular spacetimes including Schwarzschild
spacetime when one uses M${\o}$ller and Einstein prescriptions.

In spite of the efforts made during last nine decades, the problem
of localization of energy is still without a definite answer in GR.
Thus, it seems to be justified to explore this problem in the
framework of some other theories. Many authors believed that a
tetrad theory should describe more than a pure gravitation field. In
fact, M${\o}$ller [34] considered this possibility in his earlier
attempt to modify GR. Mikhail et al. [35] re-defined the M${\o}$ller
energy-momentum complex in tetrad theory. de Andrade et al. [36]
considered gravitational energy-momentum density in teleparallel
gravity. Maluf, J.W. et al. [37] explored energy and angular
momentum of the gravitational field in the teleparallel geometry.
Blagojevic, M. and Vasilic, M. [38] discussed conservation laws in
the teleparallel theory of gravity. Sakane, E. and Kawai, T. [39]
found energy-momentum and angular momentum carried by gravitational
waves in extended new GR.

Some authors [35,40,41] argued that this problem of energy
definition might be settled down in the context of teleparallel
theory (TPT) of gravitation. They showed that energy-momentum can
also be localized in the framework of this theory. It has been shown
that the results of GR and TPT agree with each other for particular
spacetimes. M${\o}$ller showed that a tetrad description of a
gravitational field equation allows a more satisfactory treatment of
the energy-momentum complex than does GR. Vargas [41] found that the
total energy of the closed Friedmann-Robertson-Walker (FRW)
spacetime is zero by using teleparallel version of Einstein and
Landau-Lifshitz complexes. This agrees with the result obtained by
Rosen [42] in GR. Recently, Yu-Xiao et al. [43] derived the
conservation laws of energy-momentum in TPG and found the
energy-momentum of the universe with the help of these laws. They
showed that the energy-momentum four-vector vanishes both in
spherical as well as in Cartesian coordinates. Also, Nester et al.
[44] explored the energy of homogenous cosmologies and discussed the
results.

In recent papers [45,46], we have obtained the TP versions of
Lewis-Papapetrou, Friedmann models and the stationary axisymmetric
solutions of the Einstein-Maxwell field equations. We have also
found the energy-momentum distribution of the Lewis-Papapetrou
spacetime [47] and the stationary axisymmetric solutions of the
Einstein-Maxwell equations by using M${\o}$ller prescription.
Further, we have evaluated the energy-momentum distribution of
static axially symmetric spacetimes [48] by using four different
prescriptions, namely, Einstein, Landau-Lifshitz, Bergmann and
M${\o}$ller. This is shown that the results of TPT do not agree with
those available in the context of GR for this particular spacetime.
In this paper, we extend this idea to evaluate the energy-momentum
density components both in GR and TPT for FRW metric. It is shown
that the results for both the theories turn out be consistent
partially.

The scheme of this paper is as follows. In section \textbf{2}, we
shall give a brief overview of the theory of teleparallel gravity.
Section \textbf{3} explains the formulation to evaluate
energy-momentum. Section \textbf{4} is devoted to determine the
energy-momentum components for the Friedmann models both in GR  and
TPT. The last section will furnish the summary and discussion of the
results obtained.

\section{Theory of Teleparallel Gravity }

The theory of teleparallel gravity is described by the
Weitzenb$\ddot{o}$ck connection given by [49]
\begin{eqnarray}
{\Gamma^\theta}_{\mu\nu}={{h_a}^\theta}\partial_\nu{h^a}_\mu,
\end{eqnarray}
where the non-trivial tetrad ${h^a}_\mu$ and its inverse field
${h_a}^\nu$ satisfy the relations
\begin{eqnarray}
{h^a}_\mu{h_a}^\nu={\delta_\mu}^\nu; \quad\
{h^a}_\mu{h_b}^\mu={\delta^a}_b.
\end{eqnarray}
In this paper the Latin alphabet $(a,b,c,...=0,1,2,3)$ will be used
to denote the tangent space indices and the Greek alphabet
$(\mu,\nu,\rho,...=0,1,2,3)$ to denote the spacetime indices. The
Riemannian metric in TPT arises as a by product [50] of the tetrad
field given by
\begin{equation}
g_{\mu\nu}=\eta_{ab}{h^a}_\mu{h^b}_\nu,
\end{equation}
where $\eta_{ab}$ is the Minkowski spacetime such that
$\eta_{ab}=diag(+1,-1,-1,-1)$. In TPT, the gravitation is attributed
to torsion [51], which plays the role of force here. For the
Weitzenb$\ddot{o}$ck spacetime, the torsion is defined as [52]
\begin{equation}
{T^\theta}_{\mu\nu}={\Gamma^\theta}_{\nu\mu}-{\Gamma^\theta}_{\mu\nu},
\end{equation}
which is antisymmetric in nature. Due to the requirement of absolute
parallelism, the curvature of the Weitzenb$\ddot{o}$ck connection
vanishes identically. The Weitzenb$\ddot{o}$ck connection also
satisfies the relation given by
\begin{equation}
{{\Gamma^{0}}^\theta}_{\mu\nu}={\Gamma^\theta}_{\mu\nu}
-{K^{\theta}}_{\mu\nu},
\end{equation}
where
\begin{equation}
{K^\theta}_{\mu\nu}=\frac{1}{2}[{{T_\mu}^\theta}_\nu+{{T_\nu}^
\theta}_\mu-{T^\theta}_{\mu\nu}]
\end{equation}
is the {\bf contortion tensor} and ${{\Gamma^{0}}^\theta}_{\mu\nu}
$ are the Christoffel symbols.

M${\o}$ller [34] poineered the teleparallel approach to
energy-momentum in 1961. Later, Mikhail et al. [35] defined their
super-potential associated with M${\o}$ller's tetrad theory
formulation. This is given as
\begin{equation}
{U_\mu}^{\nu\beta}=\frac{\sqrt{-g}}{16\pi}P_{\chi\rho\sigma}^{\tau\nu\beta}
[{\Phi^\rho}g^{\sigma\chi} g_{\mu\tau}-\lambda g_{\tau\mu}
K^{\chi\rho\sigma}-g_{\mu\tau}(1-2\lambda) K^{\sigma\rho\chi}],
\end{equation}
where
\begin{equation}
P_{\chi\rho\sigma}^{\tau\nu\beta}= {\delta_\chi}^{\tau}
g_{\rho\sigma}^{\nu\beta}+{\delta_\rho}^{\tau}
g_{\sigma\chi}^{\nu\beta}-{\delta_\sigma}^{\tau}
g_{\chi\rho}^{\nu\beta},
\end{equation}
while $ g_{\rho\sigma}^{\nu\beta}$ is a tensor quantity and is
defined by
\begin{equation}
g_{\rho\sigma}^{\nu\beta}={\delta_\rho}^{\nu}{\delta_\sigma}^{\beta}
-{\delta_\sigma}^{\nu}{\delta_\rho}^{\beta}.
\end{equation}
$K^{\sigma\rho\chi}$ is contortion tensor as given by Eq.(6), $g$ is
the determinant of the metric tensor $g_{\mu\nu}$, $\lambda$ is free
dimensionless coupling constant of teleparallel gravity, $\kappa$ is
the Einstein constant and $\Phi_\mu$ is the basic vector field given
by
\begin{equation}
\Phi_\mu={T^\nu}_{\nu\mu}.
\end{equation}
The energy-momentum density is defined as [35]
\begin{equation}
\Xi_\mu^\nu= U_\mu^{\nu\rho},_\rho,
\end{equation}
where comma means ordinary differentiation. The momentum 4-vector
can be expressed as
\begin{equation}
P_\mu ={\int}_\Sigma {\Xi_\mu^0} dxdydz,
\end{equation}
where  $P_0$ gives the energy and $P_1$, $P_2$, and $P_3$  are the
momentum components. The integration is taken over the hyper-surface
element $\Sigma$, which is described by $x^0=t=constant$. The energy
may be given in the form of surface integral using Gauss's theorem
as
\begin{equation}
E=\lim_{r \rightarrow \infty} {\int}_{{r=constant}}
{U_0}^{0\rho}u_\rho dS,
\end{equation}
where $u_\rho$ is the unit three-vector normal to the surface
element $dS$.

\section{M${\o}$ller Energy-Momentum Complex}

In 1958, M${\o}$ller presented a new \emph{pseudotensor} description
of energy-momentum for gravitating systems with an interesting
property: namely, the energy value is independent of the choice of
spatial coordinates. This is given as [8]
\begin{equation}
M_\mu^\nu= \frac{1}{8\pi}Q_\mu^{\nu\rho},_{\rho},
\end{equation}
where
\begin{equation}
Q_\mu^{\nu\rho}=\sqrt{-g}(g_{\mu\sigma,\tau}-g_{\mu\tau,\sigma})g^
{\nu\tau}g^{\sigma\rho},
\end{equation}
which is anti-symmetric in $\nu\rho$. Here $M_0^0$ is the energy
density and $ M_\mu^0~(\mu=1,2,3)$ are the momentum density
components. This satisfies the following local conservation laws
\begin{equation}
\frac{\partial {M_\mu^\nu}}{\partial x^\nu}=0,
\end{equation}
which contains contributions from the matter, gravitational and
non-gravitational fields. The momentum 4-vector of M${\o}$ller
prescription is defined as
\begin{equation}
P_\mu ={\int}{\int}{\int} {M_\mu^0} dxdydz.
\end{equation}
By using Gauss's law one can transform the last relation as
\begin{equation}
P_\mu =\frac{1}{8\pi}{\int}{\int}{Q_\mu^{0\rho}u_\rho} dS,
\end{equation}
where $u_\rho$ is the unit three-vector normal to the surface
element $dS$.

\section{Energy-momentum of the Friedmann Models}

\subsection{Energy in General Relativity}

The Friedmann models of the universe are defined by the metric [53]
\begin{equation}
ds^2=dt^2-a^2(t)[d\chi^2+{f_k}^2(\chi)(d\theta^2+\sin^2{\theta}d\phi^2)],
\end{equation}
where
\begin{eqnarray}
f(\chi)&=&\sinh\chi,\quad k=-1,\nonumber\\
&=&\chi,\quad\quad\quad k=0,\nonumber\\
&=&\sin\chi,\quad k=+1,
\end{eqnarray}
$\chi$ ($0\leq\chi<\infty$ for open and flat models but
$0\leq\chi<2\pi$ for closed model) is the hyper-spherical angle and
$a(t)$ is the scale parameter. The isotropic form of the above
metric is given as [54]
\begin{equation}
ds^2=dt^2-{\frac{a^2(t)}{(1+{\frac{1}{4}}k \rho^2)^2}}
[d\rho^2+\rho^2(d\theta^2+\sin^2{\theta}d\phi^2)],
\end{equation}
which can be written as
\begin{equation}
ds^2=dt^2-\frac{a^2(t)}{A^2(\rho)} (dx^2+dy^2+dz^2).
\end{equation}
Here $x=\rho\sin\theta \cos\phi,~ y=\rho\sin\theta \sin\phi,~
z=\rho\cos\theta,~A(\rho)=1+\frac{1}{4}k \rho^2$ and
$\rho=\sqrt{x^2+y^2+z^2}$. Using Eq.(22) in Eq.(15), we get the
following non-vanishing components
\begin{eqnarray}
Q_1^{01}=Q_2^{02}=Q_3^{03}=\frac{2\dot{a}(t)a^2(t)}{A^3(\rho)}.
\end{eqnarray}
Substituting these values in Eq.(14), the required energy-momentum
density components turn out to be
\begin{eqnarray}
M_0^0&=&0, \nonumber\\
M_1^0&=& -\frac{3k}{8\pi A^4}\dot{a}a^2x, \nonumber\\
M_2^0&=& -\frac{3k}{8\pi A^4}\dot{a}a^2y, \nonumber\\
M_3^0&=& -\frac{3k}{8\pi A^4}\dot{a}a^2z.
\end{eqnarray}
Here dot means differentiation with respect to $t$. It follows from
here that energy of the Friedmann models vanishes in GR. This
exactly coincides with that found by Rosen [42] using Einstein's
prescription. However, momentum is non-vanishing along $x,~y$ and
$z$ directions. It seems that this energy-momentum density violates
the usual energy conditions.

\subsection{ Energy in Teleparallel Gravity}

Here we evaluate the energy-momentum distribution of the Freidmann
models by using TP formulation as given in Eq.(7). Following the
procedure given in [50], we write the tetrad of Eq.(21) as
\begin{equation}
{h^a}_\mu=\left\lbrack\matrix { 1   &&&   0    &&&   0    &&&   0
\cr 0        &&& \frac{a(t)}{A(\rho)} &&&   0    &&&   0 \cr 0 &&&
0 &&& \frac{a(t)}{A(\rho)}  &&& 0 \cr 0        &&&   0    &&& 0
&&& \frac{a(t)}{A(\rho)}  \cr } \right\rbrack.
\end{equation}
Its inverse becomes
\begin{equation}
{h_a}^\mu=\left\lbrack\matrix { 1   &&   0    &&   0    &&   0 \cr
0        &&  \frac{A(\rho)}{a(t)} &&   0    &&   0 \cr 0 &&    0
&& \frac{A(\rho)}{a(t)} && 0 \cr 0 &&   0    && 0    &&
\frac{A(\rho)}{a(t)} \cr } \right\rbrack.
\end{equation}
Using these tetrad in Eq.(1) and then in (4), we obtain the
following non-vanishing components of the torsion tensor
\begin{eqnarray}
{T^1}_{10}&=&{T^2}_{20}={T^3}_{30}=- {T^1}_{01}=- {T^2}_{02}
=-{T^3}_{03}=-\frac{\dot{a}(t)}{a(t)}, \nonumber\\
{T^2}_{21}&=&{T^3}_{31}=-{T^2}_{12}=
-{T^3}_{13}=\frac{2kx}{4+k\rho^2}, \nonumber\\
{T^1}_{12}&=&{T^3}_{32}=-{T^1}_{21}=
-{T^3}_{23}=\frac{2ky}{4+k\rho^2}, \nonumber\\
{T^1}_{13}&=&{T^2}_{23}=-{T^1}_{31}=
-{T^2}_{32}=\frac{2kz}{4+k\rho^2}.
\end{eqnarray}
Using the above values in Eq.(10) and then multiplying with the
respective components of $g^{\mu\nu}$, we have
\begin{eqnarray}
\Phi^0&=&-\frac{3\dot{a}(t)}{a(t)},\quad~\quad
\Phi^1=-\frac{A(\rho)kx}{a^2}, \nonumber\\
\Phi^2&=&-\frac{A(\rho)ky}{a^2},\quad \Phi^3=-\frac{A(\rho)kz}{a^2}.
\end{eqnarray}
Substituting Eq.(27) in Eq.(6), we get the following non-vanishing
components of the contorsion tensor, in cotravariant form, as
\begin{eqnarray}
K^{011}&=&=K^{022}=K^{033}=-\frac{\dot{a}(t)A^2(\rho)}{a^3(t)}=-K^{101}
=-K^{202}=-K^{303}, \nonumber\\
K^{122}&=&K^{133}=-\frac{kA^3(\rho)x}{2a^4(t)}=-K^{212}=-K^{313}, \nonumber\\
K^{211}&=&K^{233}=-\frac{kA^3(\rho)y}{2a^4(t)}=-K^{121}=-K^{323}, \nonumber\\
K^{311}&=&K^{322}=-\frac{kA^3(\rho)z}{2a^4(t)}=-K^{131}=-K^{232}.
\end{eqnarray}
Replacing these values in Eq.(7), the non-zero components of the
superpotential are
\begin{eqnarray}
U_0^{01}&=&\frac{ka(t)x}{8\pi A^2(\rho)}, \quad
U_0^{02}=\frac{ka(t)y}{8\pi A^2(\rho)}
\quad U_0^{03}=\frac{ka(t)z}{8\pi A^2(\rho)},\nonumber\\
U_1^{01}&=&U_2^{02}=U_3^{03}= -\frac{\dot{a}(t)a^2(t)}{4\pi A^3}.
\end{eqnarray}
It is remarked here that the results do not depend on $\lambda$.
When we make use of Eq.(30) in Eq.(11), the energy-momentum density
components become
\begin{eqnarray}
\Xi_0^0&=&\frac{(12-k\rho^2)}{32\pi A^3}ka(\rho),\nonumber\\
\Xi_1^0&=&\frac{3k}{8\pi A^4}\dot{a}a^2x=-M_1^0, \nonumber\\
\Xi_2^0&=&\frac{3k}{8\pi A^4}\dot{a}a^2y=-M_2^0, \nonumber\\
\Xi_3^0&=&\frac{3k}{8\pi A^4}\dot{a}a^2z=-M_3^0.
\end{eqnarray}
It follows from here that energy is zero for the flat model, i.e.,
when $k=0$. On the spherical shell at the radius $\rho=2\sqrt{3}$,
we get the energy density which vanishes for the closed model. For
smaller values of $\rho$, the energy density is positive but for
larger values it is negative. The result of closed model partially
coincides with the earlier result found by Vargas [41] according to
which energy density vanishes for the closed model by using Einstein
and Landau-Lifshitz prescriptions. It is obvious from Eq.(31) that
the energy density remains always negative in the case of open
model. Further, the momentum turns out to be non-vanishing along
$x,~y,~z$ directions indicating that this teleparallel measure of
energy is not a good one.

\section{Summary and Discussion}

The problem of energy-momentum localization has been a subject of
many researchers but still remains un-resolved. Numerous attempts
have been made to explore a quantity which describes the
distribution of energy-momentum due to matter, non-gravitational
field and gravitational fields. Many  energy-momentum complexes have
been found [2-5] and the problem associated with the energy-momentum
complexes leads to the doubts about the idea of energy localization.
This problem first appeared in electromagnetism which turns out to
be a serious matter in GR due to the non-tensorial quantities. Many
researchers considered different energy-momentum complexes and
obtained encouraging results. Virbhadra et al. [17-22] explored
several spacetimes for which different energy-momentum complexes
show a high degree of consistency in giving the same and acceptable
energy-momentum distribution.

This paper is aimed to find energy and momentum of the Friedmann
models using M${\o}$ller's pseudotensor prescription in General
Relativity and a certain energy-momentum density developed from his
teleparallel formulation. We see from Eq.(24) that energy density
vanishes, which gives zero energy for all the three models in GR.
This energy exactly coincides with that already found by Rosen [42]
using Einstein gravitational pseudo-tensor. In TPT, energy becomes
zero for the flat model. For the closed model, energy vanishes on
the spherical shell at the radius $\rho=2\sqrt{3}$. This shows that
the energy of the closed FRW universe become consistent with Vargas
[36] found by using Einstein's and Landau-Lifshitz prescriptions.
Further, we note that momentum become zero for the flat model in
both GR and TPT. Moreover, our results in TPT are independent of the
teleparallel free dimensionless coupling constant $\lambda$. This
means that these results will be valid not only in the case of
teleparallel equivalent of GR, but also valid in any teleparallel
model. It is worth mentioning here that the components of the
momentum densities are exactly same with different signs both in GR
and TPT.

Finally, we remark that the gravitational energy exactly cancels out
the matter energy for the flat and closed universes only on the
spherical shell at the radius $\rho=2\sqrt{3}$. We see that the
energy density turns out to be negative for the open model and also
the momentum does not vanish along $x,~y,~z$ directions. These
indicate that M${\o}$ller's complex may not be a good measure of
energy-momentum.
\vspace{0.5cm}

{\bf Acknowledgment}

\vspace{0.5cm}

We would like to thank the Higher Education Commission Islamabad,
Pakistan for its financial support through the {\it Indigenous PhD
5000 Fellowship Program Batch-I}. Thanks are also due to the
referee's constructive comments.
\vspace{0.5cm}

{\bf \large References}

\begin{description}

\item{[1]} Misner, C.W., Thorne, K.S. and Wheeler, J.A.:
           \textit{Gravitation} (Freeman, New York,1973).

\item{[2]} Trautman, A.: \textit{Gravitation}: \textit{An introduction
           to Current Research}, ed. Witten, L. (Wiley, New York, 1962).

\item{[3]} Landau, L.D. and Lifshitz, E.M.: \textit{The Classical Theory
           of Fields}(Addison-Wesley Press, New York, 1962).

\item{[4]} Papapetrou, A.: Proc. R. Irish Acad. \textbf{A52}(1948)11.

\item{[5]} Bergmann, P.G. and Thomson, R.: Phys. Rev. \textbf{89}(1958)400.

\item{[6]} Tolman, R.C.: \textit{Relativity Thermodynamics and
           Cosmology} (Oxford University Press, Oxford, 1934).

\item{[7]} Weinberg, S.: \textit{Gravitation and Cosmology} (Wiley, New
           York, 1972).

\item{[8]} M${\o}$ller, C.: Ann. Phys. (N.Y.) \textbf{4}(1958)347.

\item{[9]} Cooperstock, F.I. and Sarracino, R.S.: J. Phys. A: Math.
           Gen. \textbf{11}(1978)877.

\item{[10]} Bondi, H.; \textit{ Proc. R. Soc. London } \textbf{A427}(1990)249.

\item{[11]} Penrose, R.: \textit{Proc. Roy. Soc. London } \textbf{A388}(1982)457;
            GR 10 Conference eds. Bertotti, B., de Felice, F. and Pascolini, A. Padova
            \textbf{1}(1983)607.

\item{[12]} Brown, J.D. and York, Jr. J.W.: Phys. Rev. \textbf{D47}(1993)1407.

\item{[13]} Hayward, S.A.: Phys. Rev. \textbf{D497}(1994)831.

\item{[14]} Szabados, L.B.: www.livingreviews.org/lrr-2004-4.

\item{[15]} Bergqvist, G.: Class. Quantum Grav. \textbf{9}(1992)1753.

\item{[16]} Bernstein, D.H. and Tod, K.P.: Phys. Rev. \textbf{D49}(1994)2808.

\item{[17]} Virbhadra, K.S.: Phys. Rev. \textbf{D41}(1990)1086; \textbf{D42}(1990)1066;
            and references therein.

\item{[18]} Virbhadra, K.S.: Phys. Rev. \textbf{D60}(1999)104041.

\item{[19]} Virbhadra, K.S.: Phys. Rev. \textbf{D42}(1990)2919.

\item{[20]} Virbhadra, K.S. and Parikh, J.C.: Phys. Lett.  \textbf{B317}(1993)312.

\item{[21]} Virbhadra, K.S. and Parikh, J.C.: Phys. Lett.  \textbf{B331}(1994)302.

\item{[22]} Rosen, N. and Virbhadra, K.S.: Gen. Relativ. Gravit.  \textbf{25}(1993)429.

\item{[23]} Aguirregabiria, J.M., Chamorro, A. and Virbhadra, K.S.: Gen. Relativ.
            Gravit. \textbf{28}(1996)1393.

\item{[24]} Xulu, S.S.: Int. J. Mod. Physic. \textbf{A15}(2000)2979;
            Mod. Phys. Lett. \textbf{A15}(2000)1151 and reference
            therein.

\item{[25]} Xulu, S.S.: Astrophys. Space Sci. \textbf{283}(2003)23.

\item{[26]} Sharif, M.: Int. J. Mod. Phys. \textbf{A17}(2002)1175;
            \textbf{A18}(2003)4361; Errata \textbf{A19}(2004)1495.
            Int. J. Mod. Phys. \textbf{D13}(2004)1019; Nouvo
            Cim. \textbf{B120}(2005)533.

\item{[27]} Chang, C.C., Nester, J.M. and Chen, C.: Phys. Rev. Lett.
            \textbf{83}(1999)1897.

\item{[28]} Chen, Chiang-Mei and Nester, J. M.: Class. Quantum Grav.
            \textbf{16}(1999)1279.

\item{[29]} Lessner, G.: Gen. Relativ. Gravit.  \textbf{28}(1996)527.

\item{[30]} Yang, I.C.: Ann. Phys. (N.Y.)  \textbf{4} (1958)347.

\item{[31]} Gad, R.M.: Astrophys. Space Sci. \textbf{295}(2004)495.

\item{[32]} Vagenas, E.C.: Int. J. Mod. Phys. \textbf{A18}(2003)5781;
            \textbf{A18}(2003)5949; Mod. Phys. Lett. \textbf{A19}(2003)213.

\item{[33]} Radinschi, I.: Mod. Phys. Lett. \textbf{A16}(2001)673.

\item{[34]} M${\o}$ller, C.: Mat. Fys. Medd. DanVid Selsk. {\bf 1}(1961)10.

\item{[35]} Mikhail, F.I., Wanas, M.I., Hindawi, A. and Lashin, E.I.: Int. J. Theo.
            Phys. \textbf{32}(1993)1627

\item{[36]} de Andrade, V.C., Guillen, L.C.T. and Pereira, J.G.:
Phys. Rev. Lett. \textbf{84}(2000)4533.

\item{[37]} Maluf, J.W., da Rocha-Neto, J.F., Toribio, T.M.L. and Castello-Branco, K.H.:
Phys. Rev. \textbf{D65}(2002)124001.

\item{[38]} Blagojevic, M. and Vasilic, M.: Phys. Rev. \textbf{D64}(2001)044010.

\item{[39]} Sakane, E. and Kawai, T.: Prog. Theor. Phys. \textbf{108}(2002)615.

\item{[40]} Nashed, G.G.L.: Phys. Rev. \textbf{D66}(2002)060415.

\item{[41]} Vargas, T.: Gen. Relativ. Gravit. \textbf{36}(2004)1255.

\item{[42]} Rosen, N.: Gen. Relativ. Gravit. \textbf{26}(1994)319.

\item{[43]} Liu, Y., Zhao, Z., Yang, J. and Duan, Y.: arXiv:0706.3245.

\item{[44]} Nester, J.M., So, L.L. and Vargas, T.: Phys. Rev.
\textbf{D78}(2008)044035.

\item{[45]} Sharif, M. and Amir, M. Jamil.: Gen. Relativ. Gravit. \textbf{38}(2006)1735.

\item{[46]} Sharif, M. and Amir, M. Jamil.: Gen. Relativ. Gravit. \textbf{39}(2007)989.

\item{[47]} Sharif, M. and Amir, M. Jamil.: Mod. Phys. Lett. \textbf{A22}(2007)425.

\item{[48]} Sharif, M. and Amir, M. Jamil.: Mod. Phys. Lett. \textbf{A}(2007, to appear).

\item{[49]} Aldrovendi, R. and Pereira, J.G.: {\it An Introduction to
            Gravitation Theory} (preprint).

\item{[50]} Hayashi, K. and Tshirafuji: Phys. Rev. {\bf D19}(1979)3524.

\item{[51]} Hehl, F.W., Lord, E.A. and Smally, L.L.: Gen. Relativ. Gravit. {\bf 13}(1981)1037.

\item{[52]} Aldrovandi and Pereira, J.G.: {\it An Introduction to
            Geometrical Physics} (World Scientific, 1995).

\item{[53]} Hawking, S.W. and Ellis, G.F.R.: \textit{The Large Scale Structure of Spacetime}
Cambridge University Press (1999)).

\item{[54]} Govender, M. and Dadhich, N.: Phys. Lett. {\bf B538}(2002)233.
\end{description}
\end{document}